\documentstyle[psfig]{l-aa}
\def\la{\;
\raise0.3ex\hbox{$<$\kern-0.75em\raise-1.1ex\hbox{$\sim$}}\; }
\def\ga{\;
\raise0.3ex\hbox{$>$\kern-0.75em\raise-1.1ex\hbox{$\sim$}}\; }
\begin{document}
\thesaurus{11(02.12.1; 02.12.3; 11.17.1; 11.17.4 Q 1718+4807)}
\title{ The D/H ratio at z=0.7 toward Q 1718+4807}
\author{Sergei A. Levshakov\inst{1}\thanks{{\it Permanent address:\/} 
Ioffe Physico-Technical Institute, 194021 St.Petersburg, Russia} 
\and Wilhelm H. Kegel\inst{2} 
\and Fumio Takahara\inst{3}}
\offprints{W. H. Kegel (kegel@astro.uni-frankfurt.de)}
\institute{
National Astronomical Observatory, Mitaka, Tokyo 181, Japan
\and
Institut f\"ur Theoretische Physik der Universit\"at Frankfurt 
am Main,
Postfach 11 19 32, 60054 Frankfurt/Main 11, Germany
\and
Department of Earth and Space Science, Faculty of Science,
Osaka University, Toyonaka, Osaka 560, Japan}
\date{Received December 32, 1997; accepted December 32, 1997}
\maketitle
\markboth{S.A. Levshakov et al.: D/H at z = 0.7 toward Q 1718+4807}{ }
\begin{abstract}
The apparent discrepancy between low and high D abundances
derived from QSO spectra may be caused by
spatial correlations in the stochastic velocity field.
If one accounts for such correlations,
one finds good agreement between 
different observations
and the theoretical predictions for standard 
big bang nucleosynthesis (SBBN).
In particular, we show
that the H+D Ly$\alpha$ profile observed at 
$z_{\rm a} = 0.7$ toward
Q~1718+4807 is compatible with 
$4.1\,10^{-5} \la {\rm D/H} \la 4.7\,10^{-5}$.
This result 
is consistent with our previous D/H determination for
the $z_{\rm a} = 2.504$ system toward Q~1009+2956 and, thus, 
supports SBBN. 
\end{abstract}

\section{Introduction}

From recent HST observations of a low-redshift
($z_{\rm a} = 0.7$) absorption-line system toward the quasar Q~1718+4807
($z_e = 1.084$) Webb {\it et al.} (1997a,b) deduced D/H = $1.8-3.1\,10^{-4}$. 
This ratio is
significantly higher than that derived from other quasar spectra at
$z_{\rm a} = 2.504$ [D/H = $1.8-3.5\,10^{-5}$ by Tytler \& Burles, 1996;
D/H = $2.9-4.6\,10^{-5}$ by Levshakov, Kegel \& Takahara, 
1997 (LKT, hereinafter)],
and at $z_{\rm a} = 3.572$ [D/H = $1.7-2.9\,10^{-5}$ by Tytler {\it et al.},
1996; D/H $ > 4\,10^{-5}$ by Songaila {\it et al.}, 1997].

The apparent spread of the D/H values leads some authors to assume
fluctuations in the baryon-to-photon ratio at the epoch of BBN
(see e.g. Webb {\it et al.}, and references cited therein).
On the other hand,
according to the basic idea of homogeneity and isotropy of big bang theory
the {\it primordial} deuterium abundance should not vary
in space. One can only expect that the D/H ratio 
decreases with cosmic time 
due to conversion of D into $^3$He and heavier elements in stars.
To check whether big bang nucleosynthesis has occurred
homogeneously or not,
precise measurements
of absolute values of D/H at high redshift are extremely
important.
The fundamental character of this cosmological test requires 
an unambiguous interpretation of spectral observations.

It is well known, however, that the physical parameters  derived from 
spectral data depend on the assumptions made with respect to
the line broadening mechanism. For intergalactic
absorption lines a `non-thermal broadening' is usually assumed
to be caused by large scale motions of the absorbing gas.
The commonly used microturbulent approach disregards all correlations
of the velocity field, implying
a symmetrical (Gaussian)
distribution of the velocity components parallel to the line of
sight and a symmetrical line profile. 
\begin{figure}
\vspace{2.8cm}
\hspace{0cm}\psfig{figure=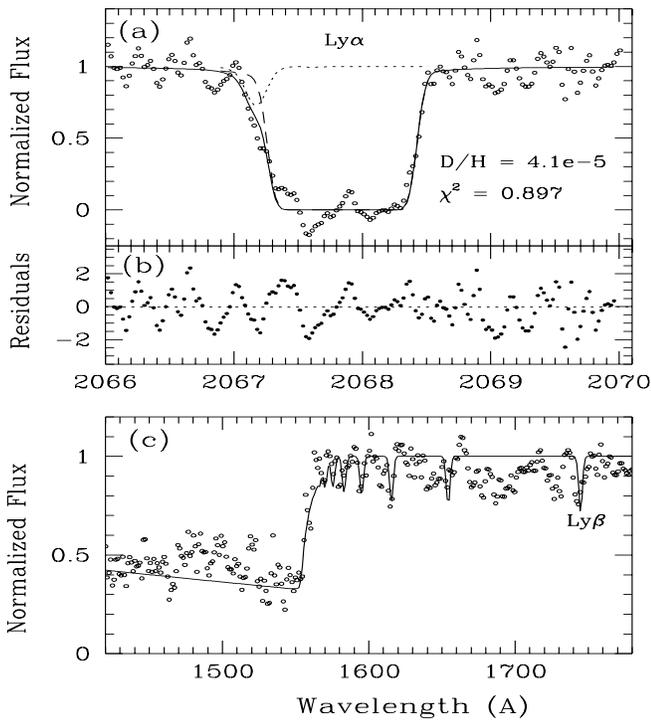,height=7.0cm,width=8.0cm}
\vspace{-0.1cm}
\caption[]{ 
Observations and RMC fits for Q~1718+4807.\, \,  
{\bf a} HST/GHRS data
(open circles) and calculated profiles for 
H\,{\sc i} (dashed curve), 
D\,{\sc i} (dotted curve)
and H+D (solid curve). The latter correspond to model (e) in Table~1.
The spectral resolution is 0.1 \AA\ (FWHM). 
{\bf b} Residuals $\epsilon$ in units of $\sigma_{\rm noise}$ (see text).
{\bf c} IUE spectrum (open circles) and fit (solid curve). The spectral
resolution is 2.95 \AA\ (FWHM)
}
\end{figure}

Actually, any turbulent flow exhibits an immanent structure in which the
velocities in neighboring volume elements 
are correlated with each other.
Different aspects
of the line formation processes in correlated turbulent media have been
discussed recently in a series of papers by Levshakov \& Kegel 
(1997, LK hereinafter),
Levshakov, Kegel \& Mazets 
(1997, LKM hereinafter), and by LKT.
\begin{figure}
\vspace{2.8cm}
\hspace{0cm}\psfig{figure=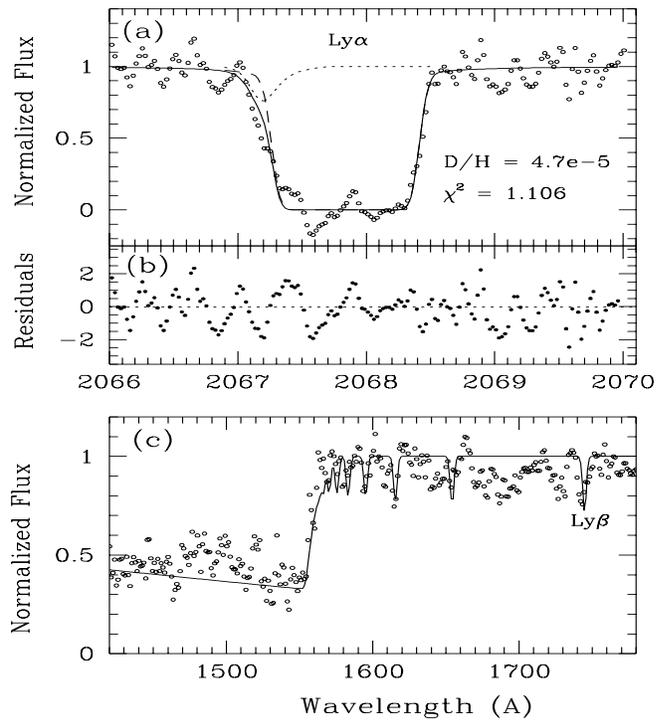,height=7.0cm,width=8.0cm}
\vspace{-0.1cm}
\caption[]{ 
As Fig. 1 but for model (c) in Table~1 }
\end{figure}

Once the statistical properties of a turbulent cloud have been 
specified,
the way spectral lines ought to be calculated, depends on the problem
considered. Considering emission lines one is dealing with many lines
of sight and the observed intensity 
should closely correspond to the theoretical
expectation value
(see e.g. Albrecht \& Kegel 1987).
If, however, one observes the absorption spectrum in 
the light of a point-like background source, 
the actually observed line profile is determined by the velocity
distribution along the particular line of  sight.
Therefore, the intensity may 
deviate substantially from the expectation value, 
since averaging 
along one line of sight corresponds to averaging over an incomplete
sample (for details see LK and LKM). 
For large values of the
ratio of the cloud thickness $L$ to the correlation length $l$ 
the distribution function $p(v)$ for the velocity component 
parallel to the line of  sight
approaches the statistical average, which has been assumed
to be a Gaussian. For values of $L/l$ of only a few, however, $p(v)$ may
deviate substantially from a Gaussian, and is asymmetric in general.
This leads to  
a complex shape of the absorption coefficient for which
the assumption of Voigt profiles could be extremely misleading.
The actual D/H ratio may turn out to be {\it higher} 
or {\it lower} than the
value obtained from the Voigt-fitting procedure.

The present Letter is primarily aimed at the inverse problem
in the analysis of the H+D Ly$\alpha$ absorption observed by 
Webb {\it et al.} (1997a,b). The original analysis was performed in
the framework of the microturbulent model. Here we  make an attempt
to re-analyze the observational data on the basis of a more general 
mesoturbulent model. We consider a cloud with a stochastic velocity
field with finite correlation length but of homogeneous 
(H\,{\sc i}-) density and
temperature. The velocity field is characterized 
by its rms amplitude  $\sigma_{\rm t}$
and its correlation length $l$.
The model is identical to that of LKT. --  The objective is to 
investigate whether the data in question may 
also be interpreted by a lower 
D/H ratio consistent with the values found for other 
absorption systems. 

\section{Parameter estimation}

To estimate physical parameters and an appropriate velocity field
structure along the line of sight, we used a Reverse Monte Carlo [RMC]
technique (see LKT). The
algorithm requires to define 
a simulation box for the 5 physical parameters~:
N(H\,{\sc i}), D/H, $T_{\rm kin}$, 
$\sigma_{\rm t}/v_{\rm th}$, and $L/l$
(here $v_{\rm th}$ denotes the thermal width of the hydrogen lines).
 -- The continuous random function of 
the coordinate $v(s)$ is represented by
its sampled values at equal space intervals $\Delta s$, i.e. by
$\{v_1, v_2, \dots , v_k\}$, the
vector of the velocity components parallel to the line of sight
at the spatial points $s_j$.

\begin{figure}
\vspace{-0.7cm}
\hspace{0cm}\psfig{figure=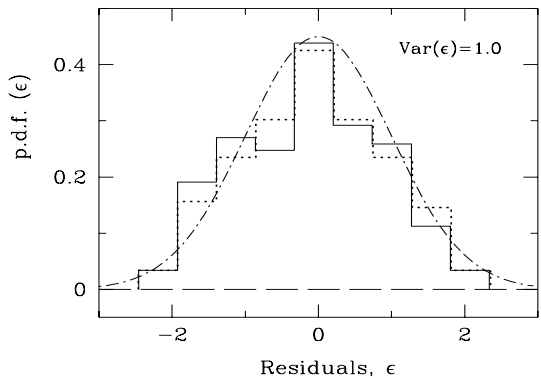,height=9.0cm,width=8.0cm}
\vspace{-3.0cm}
\caption[]{ 
Probability density functions (p.d.f.) for residuals
$\epsilon$ shown by dotted line (corresponds to Fig.1{\bf b}) and
by solid line (corresponds to Fig.2{\bf b}) and their fit by a Gaussian
with mean $E(\epsilon) = 0$ and variance $Var(\epsilon) = 1$
}
\end{figure}

In the present study we adopted for
the physical parameters the following boundaries~:

\smallskip\noindent
According to Webb {\it et al.}, N(H\,{\sc i})
is well restricted within the
range from 1.70$\,10^{17}$ cm$^{-2}$ 
to 1.78$\,10^{17}$ cm$^{-2}$, as derived from
the observed Lyman-limit discontinuity.

\smallskip\noindent
For D/H we use the range from 3.0$\,10^{-5}$ 
to 5.0$\,10^{-5}$, trying to find 
a low D/H solution.

\smallskip\noindent
To restrict $T_{\rm kin}$ and 
$\sigma_{\rm t}/v_{\rm th}$ one has to assume a model for the absorbing
material. It is generally believed that absorption line systems with
N(H\,{\sc i}) $ \sim 10^{17}$ cm$^{-2}$ arise in the halos of putative
intervening galaxies.
Direct observations
of galactic halos at $z > 2$ (van Ojik {\it et al.} 1997) show that 
$\sigma_{\rm t} \simeq 40 \pm 15$ km s$^{-1}$,  
if $T_{\rm kin} \simeq 10^4$ K.
For $T_{\rm kin}$ we use the interval $10^4 - 2\,10^4$ K, and, thus,
$\sigma_{\rm t}/v_{\rm th}$ may range within 1.3 -- 4.3 .

\smallskip\noindent
For $L/l \gg 1$ the meso- and microturbulent profiles tend to be identical
(see LK). 
Webb {\it et al.} thoroughly investigated
different microturbulent models.
Therefore we consider only moderate $L/l$ ratios in the range 1.0 -- 5.0.

Following Webb {\it et al.} (1997b), we exclude the Si\,{\sc iii} line
from our analysis of the H+D Ly$\alpha$ profile. 
As shown by Vidal-Madjar {\it et al.} (1996), `deducting lines
of sight velocity structure for D/H evaluations from ionized species
could be extremely misleading'. 
But we fix $z_{\rm a}$(Si\,{\sc iii}) = 0.701024
as a more or less arbitrary reference radial 
velocity at which $v_j = 0$.

Having specified the parameter space, we 
construct the objective function (LKT) :
\begin{equation}
{\cal L} \equiv \chi^2 = \frac{1}{\nu} \sum^{m}_{i=1}
\left[ \frac{I(\lambda_i) - r(\lambda_i)}{\sigma_{\rm noise}} \right]^2\, ,
\label{eq:E1}
\end{equation}
where $r(\lambda_i)$ is the simulated random intensity,
$I(\lambda_i)$ the observed normalized
intensity within the $i$th pixel of the line
profile, $\sigma_{\rm noise}$ the experimental error level, and
$\nu = m - n$ \,  
the degree of freedom ($m$ is the number of data points
and $n$ is the number of fitted physical parameters, 
$n = 5$ in our case).

\begin{figure}
\vspace{-0.7cm}
\hspace{0cm}\psfig{figure=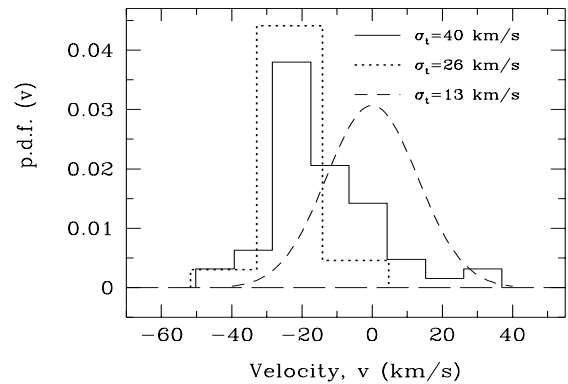,height=9.0cm,width=8.0cm}
\vspace{-3.0cm}
\caption[]{ 
Probability density functions (p.d.f.) for 
velocity components $p(v)$  for the RMC solutions shown
in Fig.1{\bf a} (dotted line histogram) and in Fig.2{\bf a}
(solid line histogram). Both are blue-shifted by about $-20$ km s$^{-1}$
with respect to $z_{\rm a}$(Si\,{\sc iii}). 
Note the asymmetric shapes. For comparison, the short dashed line shows
$p(v)$ for the microturbulent model adopted by Webb {\it et al.} 
}
\end{figure}

Here $\sigma_{\rm noise}$ is assumed to be 
constant over the entire H+D profile.
In order to estimate $\sigma_{\rm noise}$ 
we have chosen a portion of the Q~1718+4807
continuum from the left 
($\Delta \lambda_{\rm L}  = 2066.02 - 2066.90$ \AA) and the right
($\Delta \lambda_{\rm R}  = 2068.52 - 2070.01$ \AA)
hand side of the hydrogen absorption (see Fig.~1), as well as
the bottom of the Ly$\alpha$ line 
($\Delta \lambda_{\rm B} =  2067.35 - 2068.30$ \AA). 
These three regions yield 
$\sigma_{\rm L} = 0.095$, $\sigma_{\rm R} = 0.093$, and
$\sigma_{\rm B} = 0.086$, respectively. 
At $z = 0.7$ where the number density of the Ly$\alpha$ lines is far 
lower than at high redshifts, 
the experimental noise is expected to be free from the 
contamination by weak Ly$\alpha$ forest lines.
Therefore the estimated noise level corresponds to the experimental
uncertainties. 
To approximate the error level we set $\sigma_{\rm noise} = 0.09$.

Finally, our objective function contains only the blue
and the red wing of the H\,{\sc i} line ($\Delta \lambda =
2066.90 - 2067.35$ \AA\ and $\Delta \lambda =
2068.30 - 2068.52$ \AA, respectively) since they are more sensitive to
the parameter variations than the central part of the line is. 
By this we restrict
the total number of data points to $m = 30$, implying 
$\nu = 25$ and $\frac{1}{\nu}\chi^2_{\nu,\alpha} = 1.174$ 
for the credible
probability $P_\alpha = 1 - \alpha = 0.75$. 

The estimated parameters for a few  adequate RMC profile fits 
($\chi^2_{\rm min} < \chi^2_{25,0.25}$) are listed in Table~1.
The derived deuterium abundance 
$\langle {\rm D/H} \rangle \simeq 4.4\,10^{-5}$ is about 4--7 times
smaller than the limiting values of $1.8 - 3.1\,10^{-4}$ found
by Webb {\it et al.} (1997b) in the microturbulent model excluding the 
Si\,{\sc iii} line. 
To illustrate our results, we show in Figs. 1 and 2
H+D Ly$\alpha$ profiles for the two calculations
with the lowest and the highest D abundances found 
in the mesoturbulent model
(D/H = $4.111\,10^{-5}$ and $4.755\,10^{-5}$, respectively).
They are shown
by the solid curve, whereas the open circles give  
the experimental intensities $I(\lambda_i)$. 

The residuals 
$\epsilon_i = [I(\lambda_i) - r(\lambda_i)]/\sigma_{\rm noise}$
shown in Figs. 1{\bf b} and 2{\bf b}
by filled circles are normally distributed with
zero mean and unit variance (see Fig.~3).
This fact, rather trivial in case of high S/N data, becomes
crucial when dealing with spectra as noisy as the present one is, 
because of the probability to be trapped into fitting the noise
features. The good concordance of the residuals with the expected normal
distribution is here a significant argument for the validity
of the results obtained.

\begin{table}
\caption[]{Cloud parameters derived from the Ly$\alpha$
profile by the RMC method [${\rm N}_{17}$(H\,{\sc i}) column density in
units of $10^{17}$ cm$^{-2}$, 
D/H in units of $10^{-5}$,
$T_{4,\rm kin}$ kinetic temperature
in units of $10^4$ K, $\sigma_t$ turbulent velocity in km s$^{-1}$]}
\begin{flushleft}
\begin{tabular}{ccccccc}
\hline\noalign{\smallskip}
  & N$_{17}$(H\,{\sc i}) & D/H & $T_{4,\rm kin}$ & 
$\sigma_t$ & $L/l$ & $\frac{1}{\nu}\chi^2_{\rm min}$ \\ 
\noalign{\smallskip}
\noalign{\smallskip}
\hline
\noalign{\smallskip}
(a) &1.732  & 4.565 & 1.41 &  22 & 2.7 & 1.064 \\
(b) &1.739  & 4.562 & 1.60 &  18 & 3.9 & 1.002 \\
(c) &1.759  & 4.755 & 1.75 &  40 & 2.8 & 1.106 \\
(d) &1.761  & 4.555 & 1.46 &  29 & 4.3 & 1.086 \\
(e) &1.768  & 4.111 & 1.51 &  26 & 3.5 & 0.897 \\
(f) &1.771  & 4.442 & 1.76 &  23 & 3.4 & 1.162 \\
(g) &1.776  & 4.249 & 1.62 &  28 & 4.0 & 1.114 \\ 
\noalign{\smallskip}
\hline
\end{tabular}
\end{flushleft}
\end{table}

To check the $v(s)$-configurations 
estimated  by the RMC procedure,
we calculated profiles for the higher
order Lyman lines and the shape of the Lyman-limit 
discontinuity and then superposed them to the corresponding part of the 
IUE spectrum. 
The results are shown in Figs. 1{\bf c} and 2{\bf c}
where again open circles correspond to the observed intensities 
and the computed
spectra are shown by the solid curves. 
We do not find any pronounced discordance
of calculated and real spectra. 
On the other hand, the spectral resolution of 2.95 \AA\ is not 
sufficient to follow a fine velocity field structure within the
$z_{\rm a} = 0.7$ absorber. 

The derived $v(s)$-configurations are not unique. 
Table~1 demonstrates the
spread of the 
rms turbulent velocities from 
18 up to 40 km s$^{-1}$. 
It is worthwhile to emphasize once more that the projected velocity
distribution function may differ considerably from  a
Gaussian. Fig.~4 shows an example of such
distortions caused by a poor statistical sample 
(i.e. incomplete averaging)
of the velocity field 
distributions for the two cases of the lowest and the 
highest D/H ratios 
from Table~1
[model (e) and (c), respectively].
Both $p(v)$ distributions are asymmetric.
This is the main reason why the absorption in 
the blue wing of the 
H\,{\sc i} Ly$\alpha$ line may be enhanced without any
additional H\,{\sc i} interloper(s). 

\section{Conclusion}

\begin{figure}
\vspace{0.3cm}
\hspace{0cm}\psfig{figure=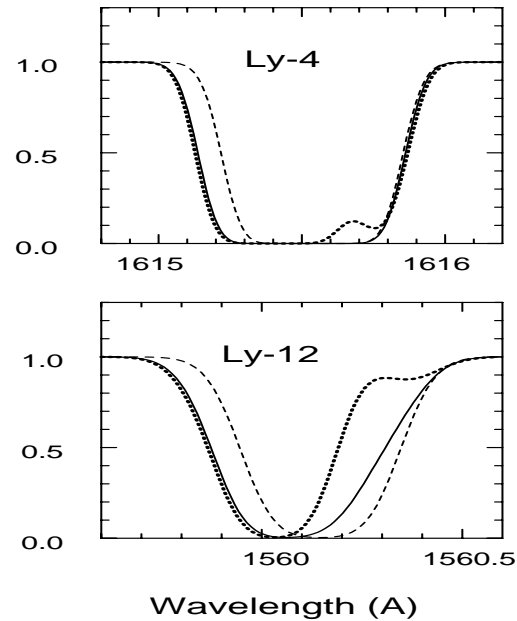,height=8.0cm,width=8.0cm}
\vspace{0.1cm}
\caption[]{ 
The H\,{\sc i} Ly-4 and Ly-12 mesoturbulent spectra 
for model (c)  
(solid curves) and (e)
(dotted curves) in Table~1. 
Short dashed curves show microturbulent
profiles calculated for the mean 
N(H\,{\sc i}) = $1.738\,10^{17}$ cm$^{-1}$
and the mean Doppler parameter $b$(H\,{\sc i}) = 25.5 km s$^{-1}$ 
of the data by Webb {\it et al.}
The spectra
are convolved with a Gaussian instrumental profile of FWHM = 0.1 \AA\ 
}
\end{figure}

We have shown that the interpretation of the HST and IUE spectra obtained 
by Webb et al. is not unique. The data can as well be modeled with a
low value of the D/H ratio if one accounts for spatial correlations
in the large scale velocity field.

The RMC results may be tested, in principle, by additional
observations of higher order Lyman lines with the same
spectral resolution as Webb {\it et al.} used for  Ly-$\alpha$.
Indeed, if $p(v)$ is
asymmetric, this will show up in the profile shapes
of the higher order Lyman lines. For the physical parameters listed in
Table~1, the effect becomes visible starting from Ly-4 
[the  Ly-$\alpha$, -$\beta$, -$\gamma$ lines are insensitive to the
asymmetry of $p(v)$ due to their high optical depth].
Fig.~5 shows simulated spectra (convolved with
a Gaussian instrumental profile of FWHM = 0.1 \AA) for Ly-4 and 
Ly-12 using
the same $p(v)$ distributions depicted in Fig.~4 -- dotted curves for
model (e) 
and solid curves for
model (c) of Table~1.
The line shapes clearly depend on
the velocity field structure and are asymmetric in general.

\vspace{1cm}

\begin{acknowledgements}The authors are grateful to John Webb 
for making available the calibrated
HST/GHRS and IUE spectra of Q~1718+4807 
and acknowledge helpful correspondence
and comments by him and Alfred Vidal-Madjar. 
This work was supported in part by the RFBR grant No. 96-02-16905a.
\end{acknowledgements}

\end{document}